\documentclass[11pt,twoside]{article}
\usepackage{graphicx,epsfig,natbib}
\usepackage{CS18}

\newcommand{\COBOLD}{{\tt CO$^5$BOLD}}

\newcommand{\LHD}{{\tt LHD}}

\newcommand{\naofe}{\ensuremath{\left[\mathrm{Na}/\mathrm{Fe}\right]}}
\newcommand{\oofe}{\ensuremath{\left[\mathrm{O}/\mathrm{Fe}\right]}}
\newcommand{\liona}{\ensuremath{\left[\mathrm{Li}/\mathrm{Na}\right]}}
\newcommand{\lioo}{\ensuremath{\left[\mathrm{Li}/\mathrm{O}\right]}}
\newcommand{\naoo}{\ensuremath{\left[\mathrm{Na}/\mathrm{O}\right]}}

\begin{document}

%

\title{Cool stars as tracers of multiple stellar populations in the Galactic globular cluster 47 Tuc}

\author{A~.Ku\v{c}inskas$^{1}$, V.~Dobrovolskas$^{1}$, P.~Bonifacio$^{2}$}

\affil{$^1$Institute of Theoretical Physics and Astronomy, Vilnius University, A.~Go\v{s}tauto 12, Vilnius LT-01108, Lithuania}
\affil{$^2$GEPI, Observatoire de Paris, CNRS, Universit\'{e} Paris Diderot, Place Jules Janssen, 92190 Meudon, France}

\begin{abstract}
Although numerous photometric and spectroscopic studies of stars in Galactic globular clusters suggest the existence of several stellar generations in the majority of clusters studied so far, kinematical properties of these stellar generations are still relatively poorly know. In this contribution we present first results of our study of kinematical properties of chemically-tagged stellar populations in the Galactic globular cluster 47~Tuc. Our analysis reveals the existence of three stellar generations in 47~Tuc that are different in their \liona\ and \naoo\ ratios. Additionally, we find that the three generations also differ in their kinematical properties: stars belonging to primordial generation are less centrally concentrated and are kinematically hotter than those in the subsequent generations.
\end{abstract}

\section{Introduction}

Results of recent photometric and spectroscopic studies of Galactic globular clusters (GGCs) suggest that most GGCs consist of at least two stellar generations that are different in their radial distribution, chemical composition, and, probably, age (see, e.g., \citealt{GCB12} for a review). An often used explanation is that this could have happened because of extended star formation that has taken place in the GGCs and which has led to the enrichment of the second generation stars with light chemical elements (such as such as Li, Na, Al) that were synthesized in the interiors of first generation stars. Under such scenario, most likely culprits responsible for the enrichment could have been either fast-rotating massive main-sequence stars \citep[e.g.,][]{CCD13}, or intermediate mass AGB stars (e.g., \citealt{DDC12}; note, however, that other explanations that do not involve the existence of multiple stellar generations have been proposed as well; see, e.g., \citealt[][]{MPL09,MZ12,BLM13,DVD14,SC14}).

$N$-body simulations predict that different stellar generations in the GGCs should differ in their spatial distributions and kinematical properties \citep{B10,B11,VMD13}. Indeed, there is plenty of evidence from recent photometric and spectroscopic studies indicating that second generation stars are more centrally concentrated \citep{GCB12}. However, kinematical properties of these different stellar generations are still largely unknown. In so far the only study of chemically-tagged stellar populations in the GGCs, \citet{BBC12} have found no statistically significant relation between the abundance of sodium and either the velocity dispersion or systemic rotation of the cluster stars. Only three clusters in their sample showed hints of dependence of radial velocity dispersion on the sodium abundance (NGC~6388, NGC~6441, and NGC~2808), although statistical significance of these trends was low. Therefore, the question whether the different stellar generations that have been observed in many GGCs do indeed differ in their kinematical properties still remains to be answered.

In this study we investigated a sample of 101 chemically tagged main-sequence turn-off (TO) stars in the globular cluster 47~Tuc, by focusing our attention on the possible relations between the chemical composition and spatial \& kinematical properties of stars that belong to different stellar generations. In what follows below we briefly summarize the first results obtained in the course of this study (for more results see \citealt[][]{KDB14} and \v{C}erniauskas et al., in preparation).

\section{Methodology}

Abundances of lithium, oxygen, and sodium that were used in our work were determined by \citet[][]{DKB14}. The abundances of all elements were measured in 101 TO stars in 47~Tuc, using for this purpose VLT~GIRAFFE spectra obtained in the gratings HR\,15N, HR\,18, and HR\,20A. In the case of oxygen and sodium, we first obtained their abundances in 1D~NLTE and then corrected them for the 3D hydrodynamical effects, by adding 3D--1D abundance corrections (i.e., the difference in the abundance of the chemical element that would be obtained from the given spectral line with the 3D hydrodynamical and 1D hydrostatic model atmospheres). Abundance corrections were computed using 3D hydrodynamical \COBOLD\ \citep{FSL12} and 1D hydrostatic \LHD\ \citep{CLS08} model atmospheres. In the case of lithium, we measured the equivalent widths of the resonance Li~I 670.8~nm line and used the obtained equivalent widths along with the formula of \citet{SBC10} to directly compute 3D~NLTE lithium abundances (for more details on the abundance determination see \citealt{DKB14}).

\begin{figure}
\centering
\includegraphics[width=8cm]{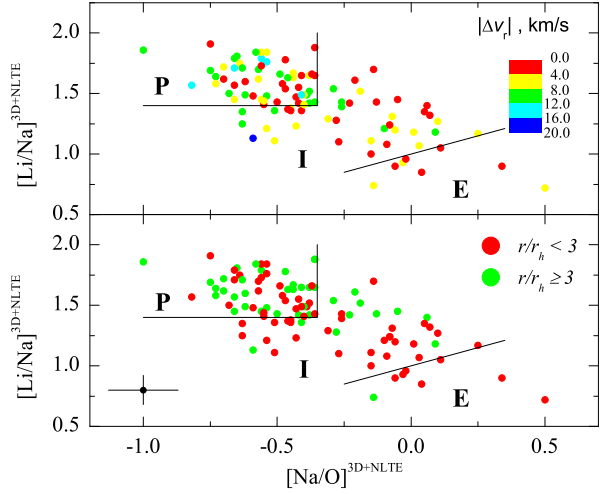}
\caption{Sample of TO stars in the $\liona-\naoo$ plane. Color-coding of individual symbols shows stars with different (a) absolute radial velocities of a given star in the reference frame of the cluster, $|\Delta v_{\rm r}|\equiv|v_{\rm rad}-\langle v_{\rm rad}\rangle^{\rm clust}|$ (top panel); and (b) distance from the cluster center, $r/r_{\rm h}$ (bottom panel; $r_{\rm h}=174^{\prime\prime}$ is the half-mass radius of 47~Tuc taken from \citealt[][]{TDG93}). Solid lines mark boundaries that were used to select stars into groups P, I, and E. Credit: Ku\v{c}inskas et al., A\&A, 568, L4 (2014), reproduced with permission \copyright ESO.
\label{fig:abund-kinem}}
\end{figure}

To study the connections between the chemical composition and kinematical properties of TO stars in 47~Tuc, stars in our sample were divided into three groups according to their position in the $\liona-\naoo$ plane: (a) primordial, P, with $\liona>1.4$ and $\naoo<-0.35$; (b) extreme, E, with $\liona < 0.6 \times \lioo +1.0$; and (c) intermediate, I, which included all remaining stars having intermediate abundances (Fig.~\ref{fig:abund-kinem}). The groups P, I, and E selected in our work approximately coincide with the three stellar generations studied in 47~Tuc earlier by \citet[][]{CBG09a} and \citet[][]{CPJ14}, where stars were assigned to different groups according to their $\oofe$ and $\naofe$ ratios. Let us note, however, that in making our selection we also tried to take into account kinematical information seen in Fig.~\ref{fig:abund-kinem}, in such a way that the group P would include nearly all Li-rich and Na-poor (O-rich) stars with the highest absolute radial velocities, $|\Delta v_{\rm r}|\equiv|v_{\rm rad}-\langle v_{\rm rad}\rangle^{\rm clust}|>8$~km/s (here $v_{\rm rad}$ is radial velocity of the individual star and $\langle v_{\rm rad}\rangle^{\rm clust}$ is the mean radial velocity of the cluster). This has ensured a better homogeneity of stars in all three groups in terms of their chemical \textit{and} kinematical properties.

\begin{figure}
\centering
\includegraphics[width=8cm]{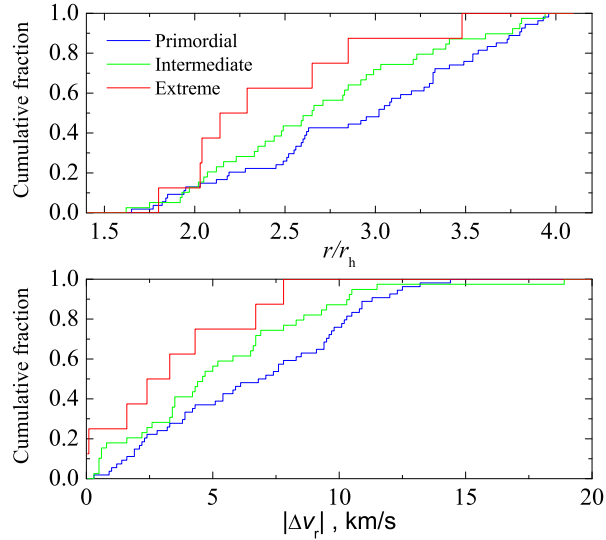}
\caption{Cumulative number fraction of TO stars in the groups P, I, and E as a function of (a) distance from the cluster center (top panel); and (b) absolute radial velocity, $|\Delta v_{\rm r}|$ (bottom panel). Credit: Ku\v{c}inskas et al., A\&A, 568, L4 (2014), reproduced with permission \copyright ESO. \label{fig:cumul-distr}}
\end{figure}

\section{Results}

First hints that stars in group P may be less centrally concentrated than those in groups I or E are seen in Fig.~\ref{fig:abund-kinem} (bottom panel). The three groups may be different in their kinematical properties, too. Indications for this are also seen in Fig.~\ref{fig:abund-kinem} (top panel), suggesting that stars in group P may be characterized by larger absolute radial velocities, $|\Delta v_{\rm r}|$, than those in groups I and E.

To verify whether these differences may be statistically significant, we carried out Kolmogorov -- Smirnov (K--S) test on the fractional distributions of stars in the three groups plotted versus (a) the radial distance from the cluster center and (b) absolute radial velocity (Fig.~\ref{fig:cumul-distr}). The K--S test performed on the fractional distributions of stars versus the radial distance showed that the probabilities for the groups P--I, P--E, and I--E being drawn from the same population were small and equal to $p=0.127$, $6.0 \times 10^{-7}$, and $0.003$, respectively. Similarly, K--S test performed on the fractional distributions of stars with respect to their absolute radial velocities showed that the probabilities in this case were equal to $p=0.069$, $7 \times 10^{-7}$, and $1.7 \times 10^{-4}$, respectively.

\begin{figure}
\centering
\includegraphics[width=8cm]{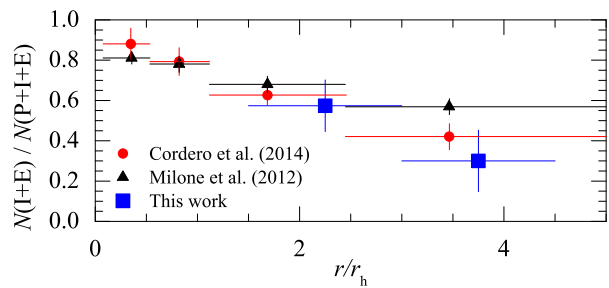}
\caption{Ratios of stars in different stellar generations in 47~Tuc, $N({\rm I+E})/N({\rm P+I+E})$, as a function of distance from the cluster center. Red circles: data from the spectroscopic study of \citet[][]{CPJ14}; black triangles: photometric results of \citet[][]{MPB12}; large blue rectangles: this work. Credit: Ku\v{c}inskas et al., A\&A, 568, L4 (2014), reproduced with permission \copyright ESO.
\label{fig:pop-ratios}}
\end{figure}

Radial distribution of the number ratios of stars in different groups, $N(I + E) / N(P + I + E)$, determined in our study agrees well with those obtained by \citet[][]{CPJ14} and \citet[][]{MPB12} from the analysis of spectroscopic and photometric data, respectively (Fig.~\ref{fig:pop-ratios}). Our results therefore confirm that 47~Tuc is not yet fully dynamically relaxed, which is in line with  theoretical predictions from $B$-body simulations \citep[see, e.g.,][]{VMD13}.

\section{Conclusions}

Our preliminary results based on the analysis of 101 TO stars in the globular cluster 47~Tuc suggest the existence of three stellar generations, which in terms of their \naoo\ abundance ratios are similar to those found in this cluster by            \citet[][]{CPJ14}. Our results also corroborate the existence of three stellar generations inferred in the analysis of \citet[][]{CGB13}, although the fractions of stars in each generation derived in the latter two studies and our work are slightly different (due to differences in the selection criteria and stellar samples used).

We find that the three stellar generations are different in terms of their radial distributions, with stars in primordial generation (group P) being less centrally concentrated than those in the subsequent generations (i.e., groups I and E). Additionally, our results indicate that stars belonging to primordial generation are kinematically hotter than those in the later generations. All these findings suggest that evolutionary history of 47~Tuc - and perhaps of other GGCs as well - may in fact be even more complex than previously thought, and therefore it awaits to be explored in the future studies based on the homogeneous analysis of larger numbers of stars.

\acknowledgments{This work was supported by grants from the Research Council of Lithuania (MIP-065/2013) and the bilateral French-Lithuanian programme ``Gilibert'' (TAP~LZ~06/2013, Research Council of Lithuania).}

\end{document}